\documentclass[a4paper,12pt,oneside]{article}
\usepackage{amsmath}
\usepackage{amsfonts,amssymb}
\usepackage{wrapfig}
\usepackage[dvips]{graphicx}
\usepackage{psfrag}
\usepackage{bm}
\usepackage{amsmath}

\makeindex
\DeclareSymbolFont{letterg}{OT1}{cmr}{m}{sl}\DeclareMathSymbol{g}{\mathalpha}{letterg}{`g}

\begin{document}

{\large
{
\begin{center}
\textbf{ELECTRON OPTIC DESIGN OF ARRAYED E-BEAM MICROCOLUMNS BASED SYSTEMS FOR WAFER
DEFECTS INSPECTION}
\end{center}
}}

\begin{center}
\textit{V.V. Kazmiruk, T.N. Savitskaja}
\end{center}

{\footnotesize
{\begin{center}
\textit{Institute of Microelectronics Technology
and High Purity Materials, Russian Academy of~Sciences}
\end{center}
}}
\vskip4pt

\begin{center}
\textbf{Abstract}
\end{center}

{\small
{In this paper is considered a~matter of~the system for wafer defect
inspection (WDIS) practical realization. Such systems are on the agenda as
the next generation and substitution for light optics and single $e$-beam
based WDISs.
}}

{\large
{
\begin{center}
\textbf{Introduction}
\end{center}
}}

At the present time an activity in the field of~e-beam micrcolumns practical
realization is growing up rapidly. The most significant progress is attained
by groups of~T.H.P.~Chang from IBM Research Center [1], P.~Kruit from Delft
Technical University [2--6] and H.S. Kim and alii [7, 8].

However, their efforts directed mainly on $e$-beam lithography
application or
just microcolumn electron optics design.

In this paper is considered a~matter of~the system for wafer defect
inspection (WDIS) practical realization. Such systems are on the agenda as
the next generation and substitution for light optics and single $e$-beam
based WDISs.

In our previous work [9] the
requirements to WDIS have been considered  as informative system
with resolution down to 2~nm. It
was shown that in the range of~10$\div $30~nm multibeam WDIS for
topographical defects inspection would be comparable in throughput with the
light optics system when number of~columns in the array is about 1000. In
the case of~the line width measuring (LWM) or surface microrelief
reconstruction can be realized resolution 2--10~nm.

Are considered aspects of~WDIS design for both application.
\vskip15pt

{\large
{
\begin{center}
\textbf{The electron optics design}
\end{center}
}}

First of~all consider the main principles of~electron optics design of~the
microcolumn. We start from simple single lens column used by many authors
[1, 8] for experiments in this field.

The electron optical components of~a~one lens column are shown schematically
in fig.\:1.

The resolution of~the microscope column is limited primarily by the
aberrations of~the objective lens.
The probe size is given by:
$$
d_{b}^{2 }= (M\cdot d_{0})^{2}+d_{d}^{2}+d_{s}^{2}+d_{c}^{2}, \eqno(1)
$$
where $M$ is the column magnification; $d_{0}$ is the virtual
source size;
$$
d_{d }= 1,5\cdot\alpha V^{(0,5)} \eqno(2)
$$
\par\pagebreak

\begin{figure}[htbp]
\centerline{\includegraphics[scale=1.4]{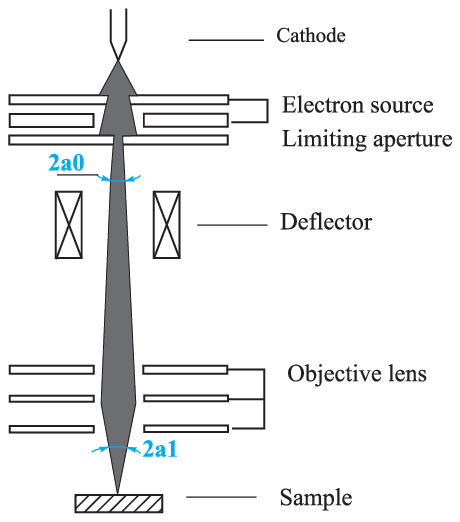}}
\label{fig1}
\end{figure}

\centerline{Fig. 1. A one lens column for $e$-beam lithography}

$\phantom{a}$
\vspace*{-1mm}

\noindent
is the diameter of~the diffraction
disk so that $d_{d}/2$ is full width half
maximum of~corresponding distribution.
$$
d_{s}= 1/2\cdot C_{s}\alpha ^{3}
$$
is the spherical
aberration disc with the spherical aberration coefficient $C_{s}$;
$$
d_{c}=C_{c}\alpha \Delta V/V
$$
 is the chromatic aberration
disk with $C_{c}$ being aberration coefficient and DV being the energy
spread of~the beam;
$$
\alpha _{0  } = \alpha \cdot M, \eqno(3)
$$
where $\alpha _{ 0 }$ is the semi convergent angle at the exit of~the
source and $\alpha $ is the semi convergent angle at the target.
Final probe current $I_{b}$ is
$$
I_{b }= p \alpha _{ 0}^{2}dI_{0}/d\Omega _{0}. \eqno(4)
$$
Here $dI_{0}/d\Omega _{0}$ is angular emission density.

We use the conventional rule (1) to estimate the chromatic
and spherical
aberration coefficients of~the objective
lens: $C_{c}$ and $C_{s }$.

Fig. 2 shows the performance of~1 keV microcolumns with two different
objective lenses [1]. The first in solid line, represents a~\textit{fixed} symmetric
einzel lens. This lens has a~200~$\mu$m bore diameter and 250~$\mu$m
spacing. The lens. operating for a~1~mm working distance in the accelerating
mode, has a~chromatic and spherical abeiration coefflcients of~approximately
2~mm and 50~mm respectively. As shown in the figure, a~probe size of~9,9~nm
can be achieved al an optimum semiconvergent angle of~$ \approx 6,3$~mrad.
Further improvement of~resolution can be achieved by optimizing the
electrodes geometry for working distance 1~mm, that allows to decrease both
spherical and chromatical coefficients to values shown by dashed lines. As a
result the resolution 8,8~nm\linebreak

\begin{figure}[htbp]
\centerline{\includegraphics[scale=0.7]{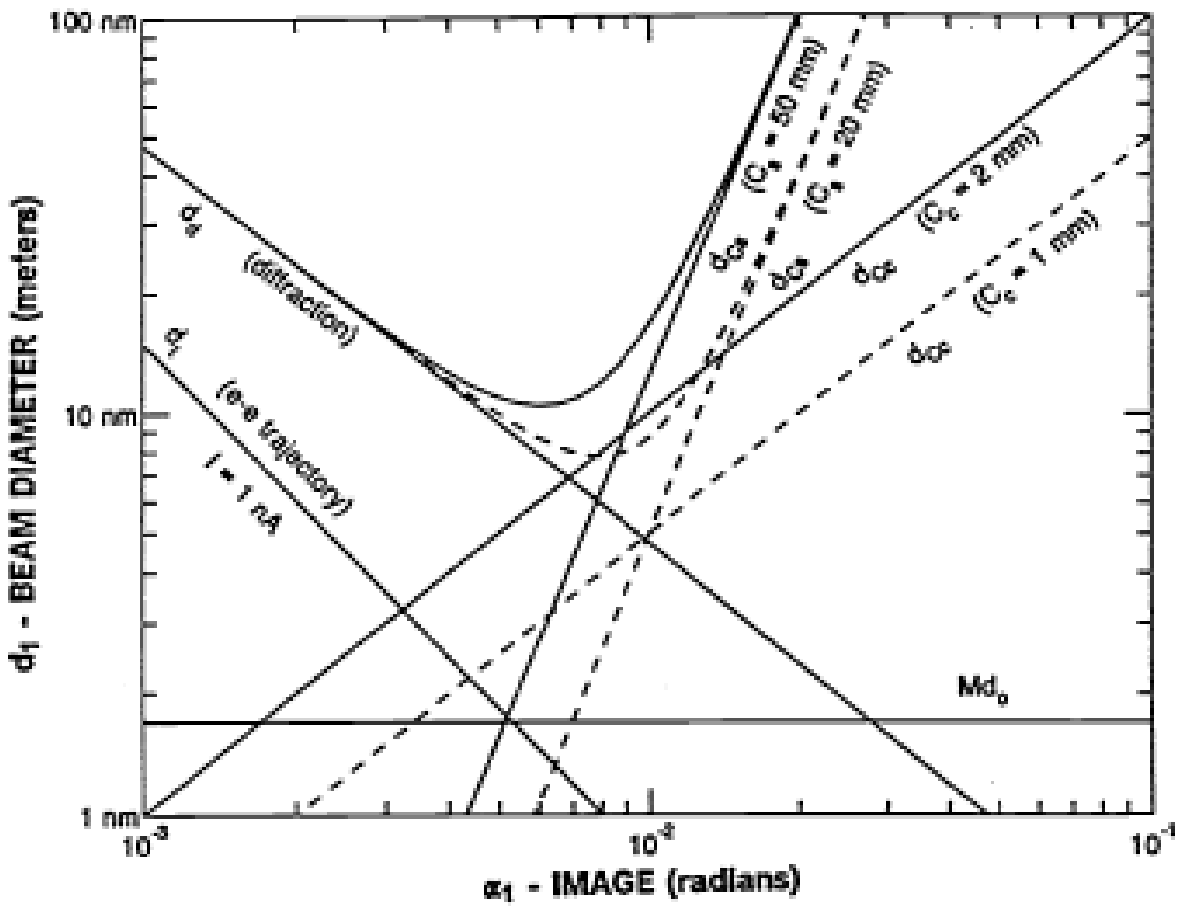}}
\label{fig2}
\end{figure}

$\phantom{a}$
\vspace*{-7mm}

\centerline{Fig. 2.}

$\phantom{a}$
\vspace*{-1mm}

\noindent
at working distance 1~mm can be achieved. These
are typical results achieved practically so far [1, 8].

It should be noted that those results achieved in transmission mode, and
working distance 1~mm is chosen to place on-axis detector between lens
and sample as it shown in fig. 3.

\begin{figure}[htbp]
\centerline{\includegraphics[scale=1.1]{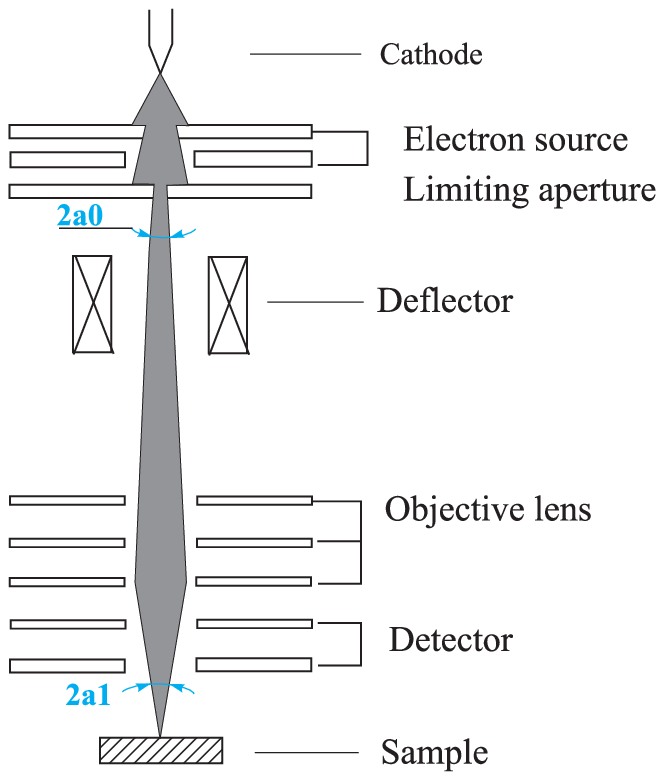}}
\label{fig3}
\end{figure}

\centerline{Fig. 3. A one lens microcolumn with on-axis detector}

$\phantom{a}$
\vspace*{-1mm}

Now consider what should be changed for improving a~resolution to 2~nm.
Is assumed that electrons energy still is 1~keV and the energy
spread $\Delta   V = 1$~eV.
In the fig.\:4 is shown an electron optical
performance of~1~keV improved
column for $C_{s}=0{,}3$~mm, $C_{c}=0{,}084$~mm.

\begin{center}
\includegraphics[scale=0.7]{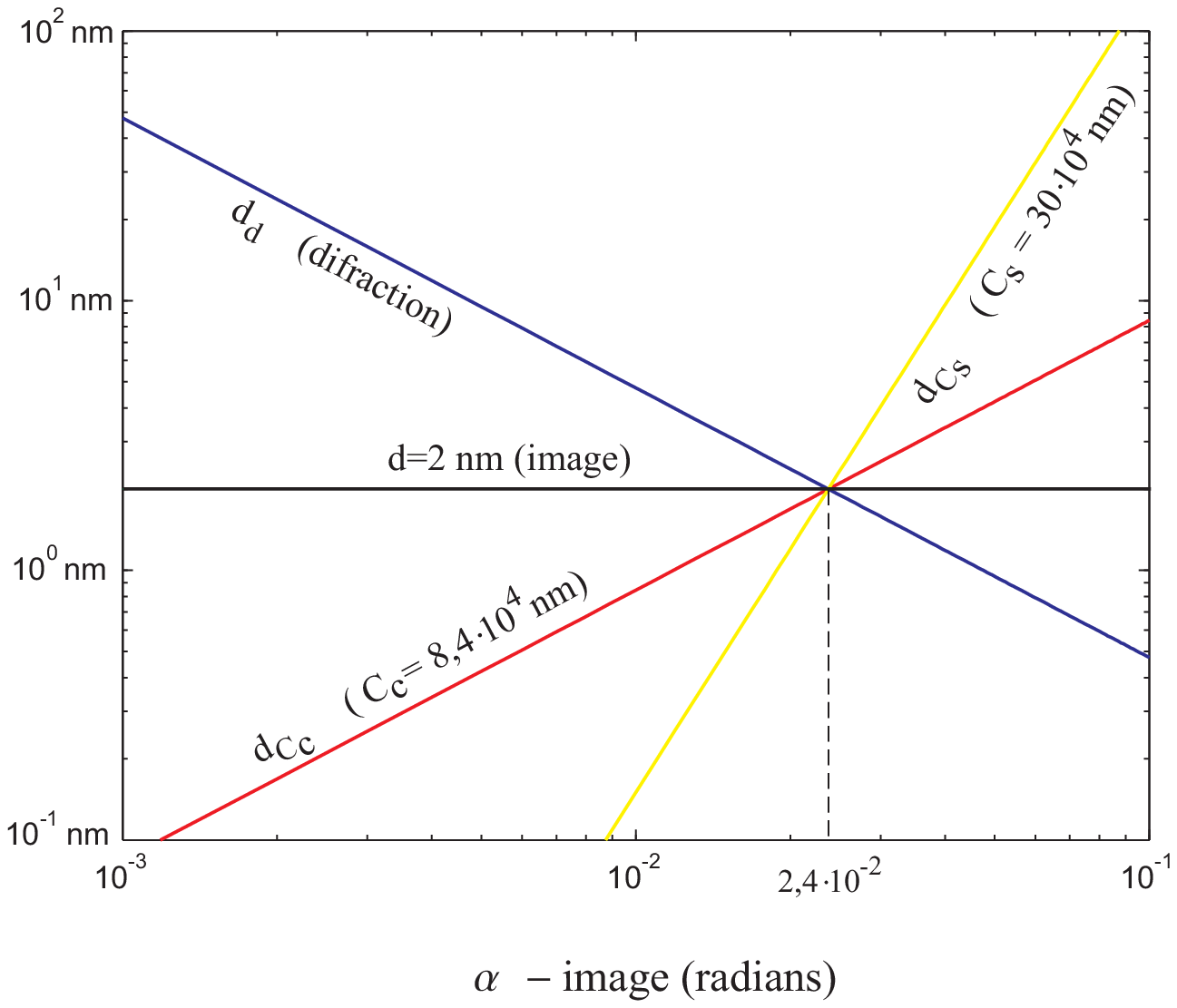}
\end{center}

\begin{center}
Fig. 4. Electron optical performance of~1~keV improved column.
$C_{s}=0,3$~mm, $C_{c}=0,084$~mm
\end{center}

It is obvious from (1) that the value $d_{b} = 2$~nm can be achieved when
each of~$d_{d}$, $d_{s}$, $d_{c}$, is less than that value. Thus,
diffraction limit becomes a~dominating factor which determines semi
convergent angle. If is chosen $\alpha \gg  2,4\cdot 10^{-2}$ radians,
then aberration coefficients $C_{s}\ll 0,3$~mm and $C_{c}\ll 0,08$~mm.
For more exact evaluation examine the residual
$$
d^2-(d_d^2+d_c^2+d_s^2)=(M\,d_0)^2
$$

\noindent
at $d_b=2$~nm and minimize $(d_d^2+d_c^2+d_s^2)$  over
$\alpha $.

Thus for given aberration coefficients $C_s$ and  $C_c$
the maximum value of~$(M\,d_0)^2$ can be calculated.

Figures 5, 6 show result for $C_{c}=0{,}04$ and $C_{c}=0{,}02$~mm.

\begin{center}
\includegraphics[scale=0.8]{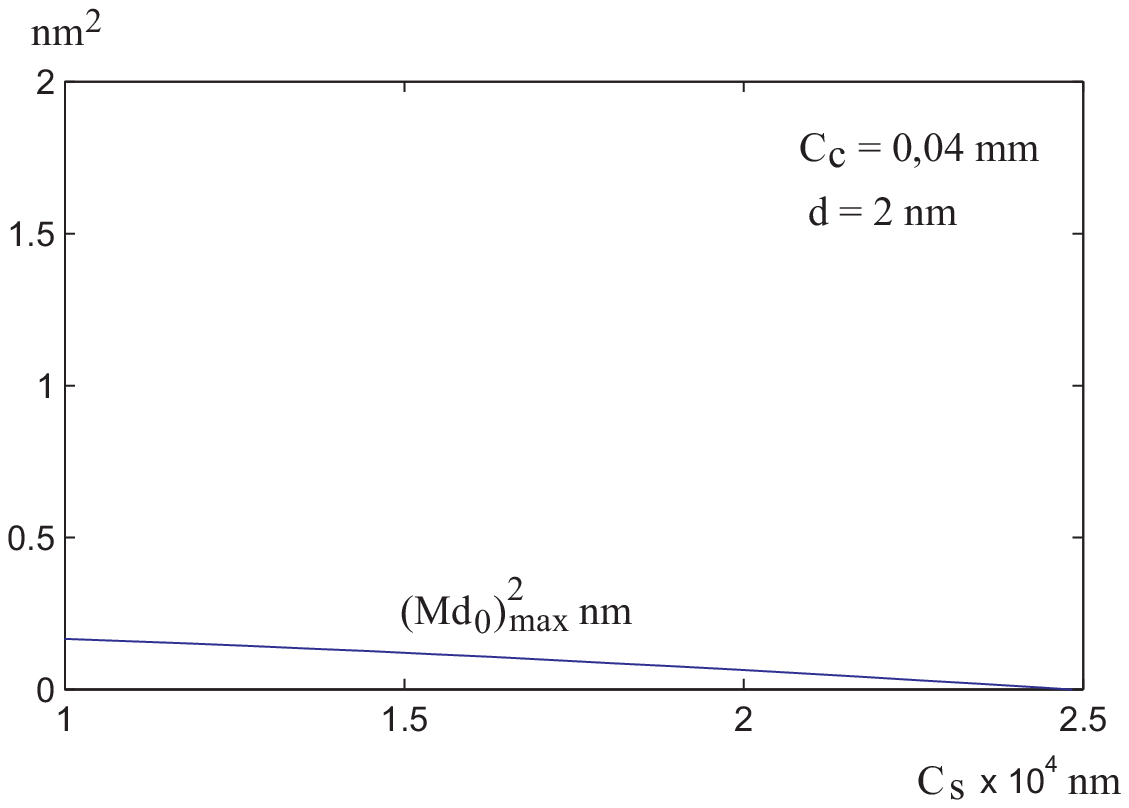}
\end{center}

\begin{center}
Fig. 5.
\end{center}

\begin{center}
\includegraphics[scale=0.8]{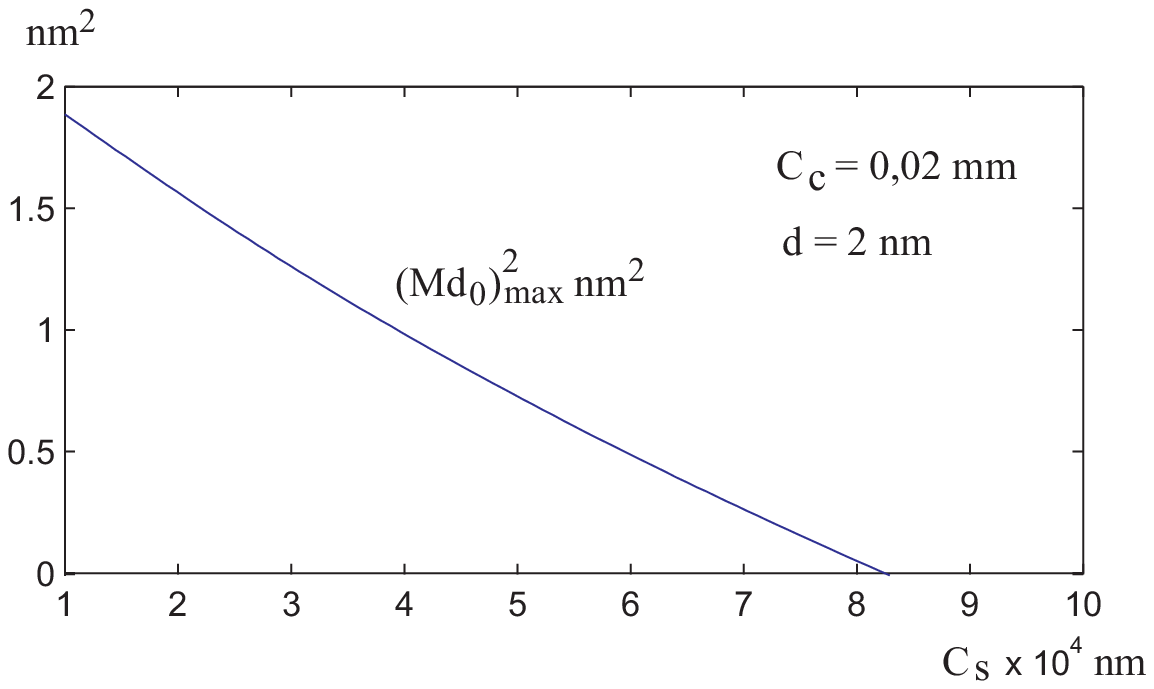}
\end{center}

\begin{center}
Fig. 6.
\end{center}

Thus, to receive the probe size 2 nm for objective
lens with the chromatic
aberration coefficient $C_{c}=0{,}04$, the spherical
aberration coefficient
$C_{s}$ needs to be lower $0{,}02$~mm and
semiconvergent angle $\alpha >2{,}6\cdot 10^{-2}$~rad.

Similarly  for $C_c=0,02$~mm $C_s\leq 0{,}08$~mm and
$\alpha>2,7\cdot10^{-2}$~rad is required.

The use another formulas for probe size diameter calculation,
for example
[2], gives no principal change to order of~$C_{s}$ and $C_{c}$ values.

Such a~way from the above analytical performance consideration follows that
for improvement of~the resolution to 2~nm it's necessary to keep semi
convergent angle more than $2,7\cdot 10^{ - 2}$~rad and radically
decrease both
chromatical and spherical aberrations coefficients.

Methods of~improvement aberrations consist in electrostatic
lens dimensional
scaling down from conventional lens. In fig.\:7 is schematically shown
spherical aberration for a~positive and a~negative lens,
illustrated with
two rays entering the lens at different radii, $r_1$ and $r_2$.
In both cases, the
intercept with the $z$-axis shifts in the negative $z$
direction for increasing
radius of~incidence.

\begin{figure}[htbp]
\centerline{\includegraphics[scale=1.5]{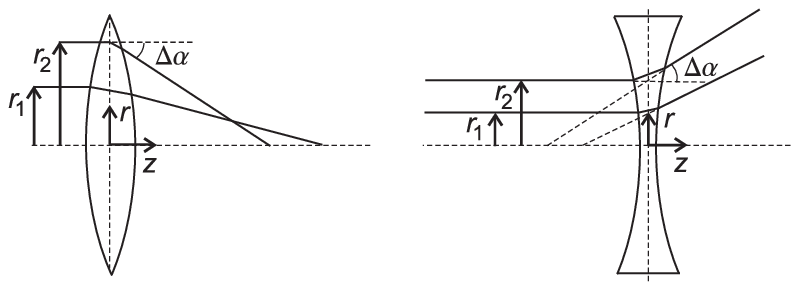}}
\label{fig4}
\end{figure}

\centerline{Fig. 7. Spherical aberration for positive and negative lens}

$\phantom{a}$
\vspace*{-1mm}

By decreasing lens bore diameter and, therefore, radius of~incidence
$r$ to a~few microns is possible to achieve even less values than
$C_{c}=0{,}02$~mm $C_{s}\leq 0{,}08$~mm.
Unfortunately, we should keep the value of~semi convergent angle $\alpha
\geq  2{,}7\cdot 10^{ - 2}$~rad, which leads the working
distance shortening.\linebreak As
working distance $\mbox{WD} = r/\alpha $, then for
incidence radius in the range
2$\div$10 $\mu$m WD should be in the range 74 $\div$370~$\mu$m. Such
short working distance does not give enough space for detector placement.

In practice $\alpha $ is even more than $2{,}7\cdot 10^{ - 2}$~rad
in order to
obtain small diffraction term.
\vskip8pt

\noindent
{
{Table 1. Two lens system performance
}}

$\phantom{a}$
\vspace*{-7mm}

\begin{table}[htbp]
\begin{tabular}
{|p{185pt}|p{185pt}|}
\hline
Probe size&
1,73 nm \\
\hline
Probe current&
1 nA \\
\hline
Magnification (specimen -- gun)&
$-2{,}31$ \\
\hline
$C_s$ gun side &
0{,}68 mm \\
\hline
$C_c$ gun side&
0{,}037 mm \\
\hline
Spherical aberration term&
0,92 nm \\
\hline
Chromatic aberration term&
0,53 nm \\
\hline
Diffraction term&
1,04 nm \\
\hline
\end{tabular}
\label{tab1}
\end{table}


\begin{figure}[htbp]
\centerline{\includegraphics[scale=1.3]{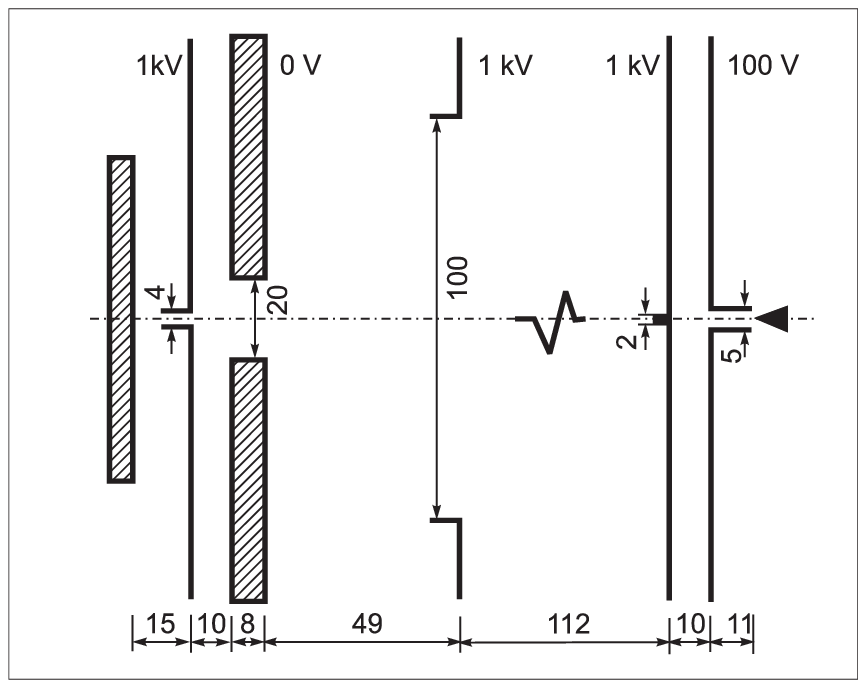}}
\label{fig5}
\end{figure}

\centerline{Fig. 8. Optimized two lens system
for $\mbox{WD} = 15$~$\mu$m [10]}

$\phantom{a}$
\vspace*{-1mm}

As an example in fig.\:8 is shown an optimized five
electrode (two lens)
system for
$\mbox{WD} = 15$~$\mu$m ($\alpha  = 0{,}13$~rad) and
object to image distance 215~$\mu$m.
In tab.\:1 is depictured the system performance.

The short WD brings a~lot others inconveniences besides mentioned above
problems with detector placement. First of~all these are vacuum
deterioration between sample and the system, risk of~wafer damaging and
small depth of~focus.
That's why the ideas about an application of~such systems are directed
mainly on lithography or average resolution low voltage SEM.
\vskip15pt

\begin{center}
{\large{\textbf{Ultra thin film foil implementation for improvement the miniature
beam system performance}}}
\end{center}

\vskip8pt

There are two promising applications of~ultra thin foils for electron
microscopy: the tunnel junction emitter and the low energy corrector. Common
for both applications is that the electron beam is sent through the thin
foil at low energy. Measurements of~mean free path for number of~metals
indicate the value about 5 nm at the energy $ \approx  5$~eV
above the Fermi
level.

First achieved by us [6] free standing foils have been 5~nm of~thickness and
later we achieved foils with thickness 4, 3 and even 2,2~nm. A substantial
part of~electrons can be transmitted through such thin film without
scattering, so film acts as an ideal energy filter.
\vskip15pt

\noindent
{\bfseries \itshape The tunnel junction emitter}

\vskip8pt

Electron field emitters are used in a~wide variety of~applications, such as:
electron microscopes, electron beam lithography machines, field emission
displays and vacuum micro electronics. Field emitters have some important
advantages over thermionic emitters: they have a~higher brightness and lower
energy spread, they can operate at ambient temperature and they have a~lower
power consumption because no heating of~a~filament is required.

Nevertheless, improvements are still desirable. For example, as it is shown
above, the spatial resolution in low voltage electron probes is limited in
part by the energy spread of~the field emitter. If it would be possible to
operate a~field emitter at low voltage, battery driven applications are in
reach (e.\,g. displays for laptop computers). The tunnel junction emitter is
expected to combine the properties of~low energy spread, high brightness,
operation at low voltage and low power consumption.

The tunnel junction emitter [2] is constructed by placing a~sharp tip within
tunneling range of~a~very thin metal foil (see fig.\:9). Between tip and
foil a~voltage larger than the work function of~the foil surface is applied.
Provided that the foil is sufficiently thin, a~fraction of~the tunneled
electrons will travel through the foil without scattering. Electrons with
sufficient forward energy to overcome the work function are emitted into the
vacuum . In this way the work function acts as a~high-pass energy filter.
Combined with the fact that the electrons originate from an atomic size
tunneling area, a~monochromatic high-brightness electron source is expected.
As for most metals the work function is of~the order of~a~few eV, the source
is operated at low voltage. Although the emitted current is only a~fraction
of~the tunnel current, the power consumption is still low because of~the low
voltage operation and because no heating is required. The emitter can be
operated at high frequency because only a~small voltage difference is needed
to switch between on and off and because the size of~the emitter, and
therefore its capacitance, can be kept small. This could be interesting for
RF applications.

As a~tip is assumed to implement very sharp tungsten tip (often called
nanotip) similar to that for STM investigations [11].
However, the experiments with clean nanotip and free standing foil as it
sketched in fig.\:9 have shown that thin film is damaging in a~short
time. That happens because of~attractive forces between tip and foil.

\begin{figure}[htbp]
\centerline{\includegraphics[scale=1.8]{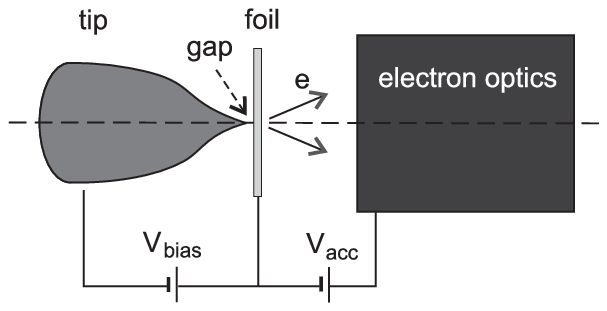}}
\label{fig6}
\end{figure}

\centerline{Fig. 9}

$\phantom{a}$
\vspace*{-1mm}

To avoid this problem was proposed another [3] configuration of~experiments
(see fig.~10).


$\phantom{a}$
\vspace*{-1mm}

\begin{figure}[htbp]
\centerline{\includegraphics[scale=0.8]{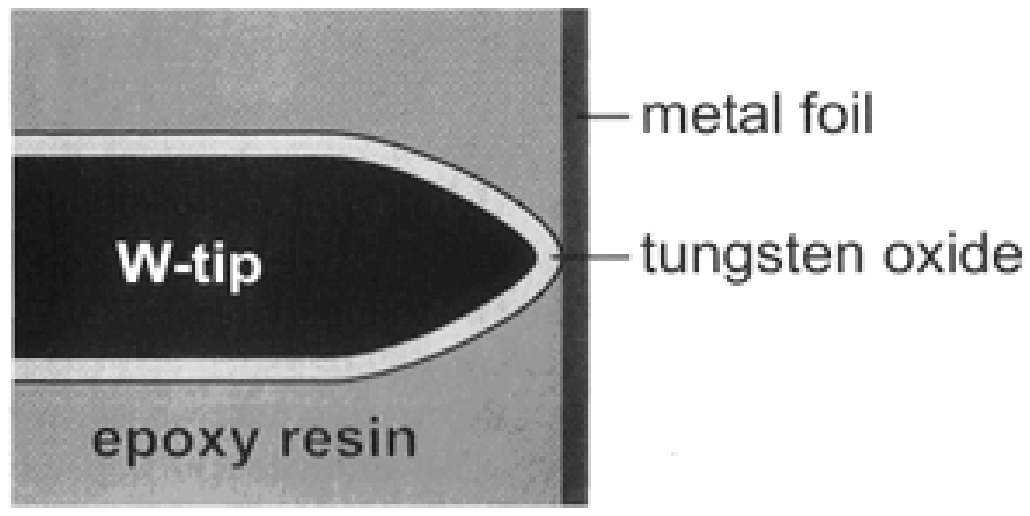}}
\label{fig6}
\end{figure}

\centerline{Fig. 10. Cross section sketch of~the device [3]}

$\phantom{a}$
\vspace*{-1mm}

\begin{figure}[htbp]
\centerline{\includegraphics[scale=0.9]{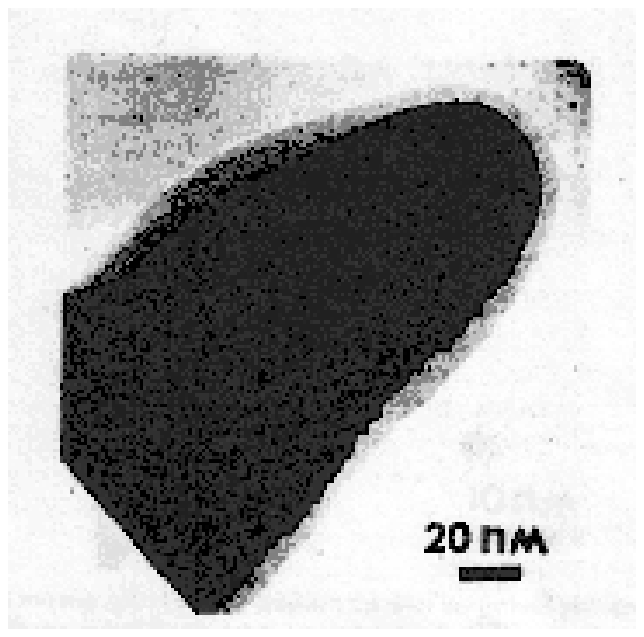}}
\label{fig7}
\end{figure}

\centerline{Fig. 11. TEM micrograph of~9V dc
etched oxide covered tungsten tip [11]}

$\phantom{a}$
\vspace*{-1mm}

As a~tip was used an oxidized tungsten tip shown at fig.\:11.

The experimental work has verified the principle of~operating of~this
emitter. However, the stability and life time are not sufficient enough and
still have to be improved. Nevertheless there is a~hope that after some
optimization of~oxide layer, choice of~proper tip material and development
of~reliable assembling technology would be possible to create a~working
device.

It should be noted one more attractive property of~solid state emitter that
it is expected to be not that critical to vacuum condition as convenient
field emitter.

\vskip15pt

\noindent
{\bfseries \itshape The low energy aberration corrector}

\vskip8pt

In the basic form corrector is sketched in fig.~12.

\begin{figure}[htbp]
\centerline{\includegraphics[scale=0.9]{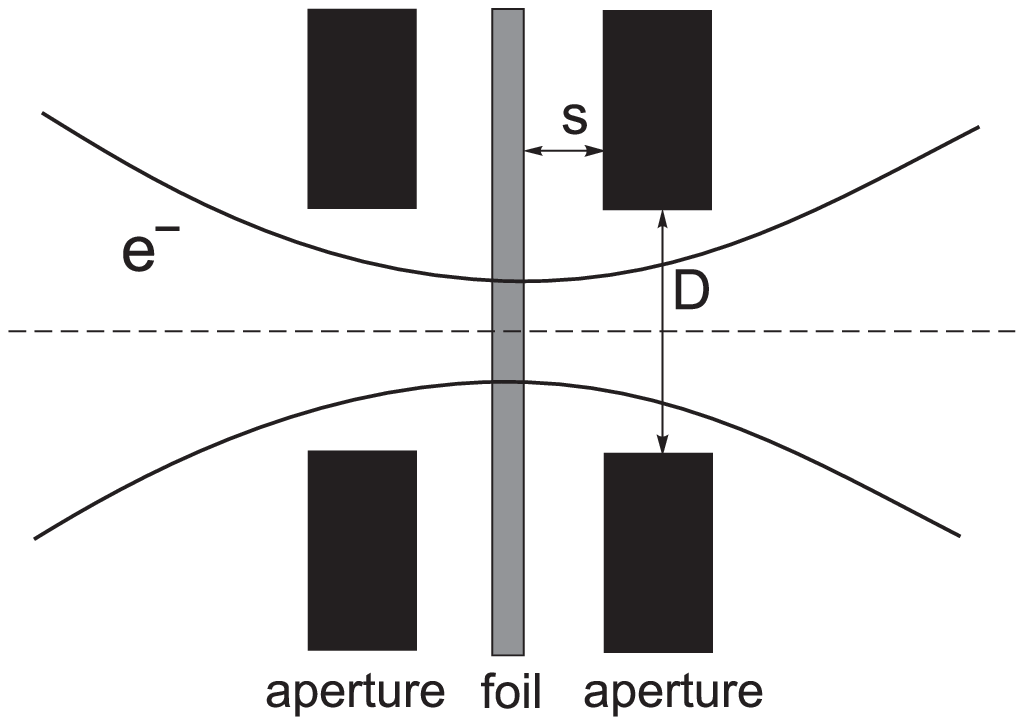}}
\label{fig8}
\end{figure}

$\phantom{a}$
\vspace*{-10mm}

\begin{center}
Fig. 12. Basic design of~the foil corrector (not to scale). D:
diameter of~the aperture; s: gap between foil and aperture [4]
\end{center}

 It consists of~a~flat free-standing foil of~nanometer size thickness with
apertures on both sides. In the low-energy foil corrector, the foil is put
on a~retarding potential, such that the electrons have almost 0~eV kinetic
energy when they enter the foil (and also when they have just left the foil
at the other side).

For use in a~SEM additional optics is necessary
to~adjust correction and to~focus the beam.
A SEM column with corrector is shown at fig.~13.

\begin{figure}[htbp]
\centerline{\includegraphics[scale=1.5]{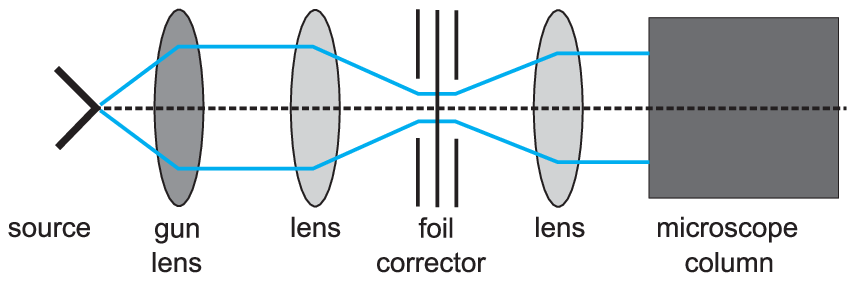}}
\label{fig9}
\end{figure}

\centerline{Fig. 13. SEM column with the foil corrector}

$\phantom{a}$
\vspace*{-1mm}

As corrector is very strong negative lens it is necessary to put focussing
lenses close to it. Because of~that reason are favorable electrostatic
lenses.

Leaving apart all details of~corrector calculations (see for details [12])
we present a~final result~--- a~calculation example on realistic set-up, e.\,g.
aberration corrected low voltage SEM (see fig.~14).

In the tab.\:2 the measures
of~the design are given  and electrode potentials
and calculation result for optimum setting are given
in the tab.\:3.

\begin{figure}[htbp]
\centerline{\includegraphics[scale=1.6]{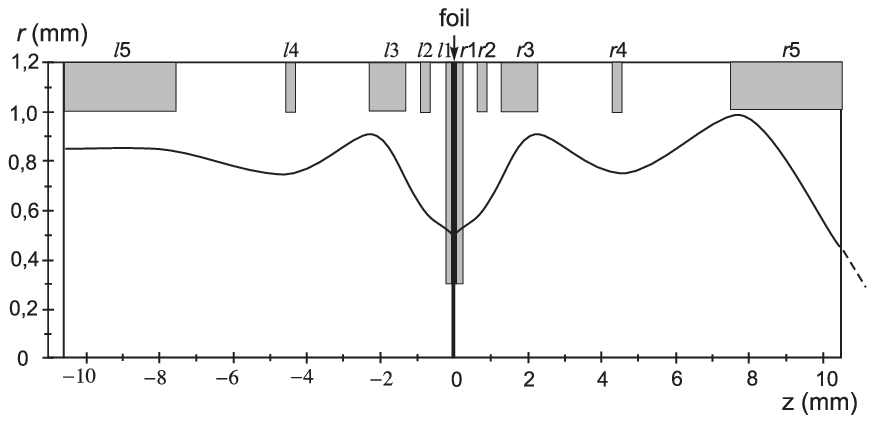}}
\label{fig10}
\end{figure}

$\phantom{a}$
\vspace*{-10mm}

\begin{center}
Fig. 14. Design of~a~low-voltage SEM column with a~low-voltage
foil corrector [5]. The design is rotationally symmetric in the $z$-axis
\end{center}

$\phantom{a}$
\vspace*{-4mm}

Above
the drawing, the numbering of~the electrodes is designated. In the drawing,
the paraxial ray for the settings in tab.\:3 is shown. For the visibility,
its radial extent is drawn 5 times larger than the maximum beam radius. Note
that this ray has started in fin object position which is far left from the
left border, thus it enters the column with a~very small, but non-zero
slope.
\vskip13pt

\noindent
{{Table 2. Measures of~the  column design in fig.\:14. The
design is mirror symmetric in the foil, only the measures for the
electrodes at the right side are listed} }

\begin{table}[htbp]
\begin{tabular}
{|p{73pt}|p{90pt}|p{100pt}|p{85pt}|}
\hline
electrode no.&
thickness (mm)&
aperture radius\par (mm)&
distance to next\par electrode (mm) \\
\hline
foil&
0&
---&
0,04 \\
\hline
$rl$&
0,20&
0,30&
0,40 \\
\hline
$r2$&
0,25&
1,00&
0,40 \\
\hline
$r3$&
1,00&
1,00&
2,00 \\
\hline
$r4$&
0,25&
1,00&
3,00 \\
\hline
$r5$&
3,00&
1,00&
--- \\
\hline
\end{tabular}
\label{tab2}
\end{table}

\pagebreak

\noindent
{{Table 3. Optical  properties calculated with aberration
integrals for a~foil potential of~respectively 0,1, 0,4 and 1,0~V.
The potentials of~the other electrodes with respect to the foil are
the same as in tab.\:2} }

\begin{table}[htbp]
\begin{tabular}
{|c|c|c|c|c|}
\hline
{optical } &
{aberration } &
 {ray tracing}&
 {electrode no.} &
 {potential (V)}
  \\
  &&&&\\[-21pt]
property&
integrals&
&&
\\
\hline
$Z_{0}$ &
$-394$~mm &
$-515$~mm  &
$l5$&5000,1
\\
\hline
$Z_{i}$ &
12,3~mm &
12,3~mm &
$l4$&8900,1
  \\
\hline
$a_{1}$ &
$-0,275$~mm$^{ - 1}$ &
$-0,272$~mm$^{ - 1}$&
$l3$&770,1
     \\
\hline
$M$ &
$-0,0204$ &
$-0,0155$ &
$l2$&2950,1
 \\
\hline
$C_{s3}$ &
$-0,68$~mm &
18~mm &
$l1$&340,1
  \\
\hline
$C_{c1}$ &
$-0,002$~mm &
$-0,83$~mm &
foil& 0,1
 \\
\hline
$C_{s5}$ &
 &
$3,2\cdot 10^{3}$~mm &
$rl$&340,1
  \\
\hline
$C_{c2}$ &
 &
$5,3\cdot10^{2}$~mm&
$r2$&2950,1
 \\
\hline
&&&
$r3$&770,1\\
\hline
&&&
$r4$&9000,1
\\
\hline
&&&
$r5$&1000,1\\
\hline
\end{tabular}
\label{tab3}
\end{table}

In fig.\:15 it is shown probe size versus the probe current divided by
transmission ratio of~current through the foil. The latter is equal to the
current incident on the foil

\begin{figure}[htbp]
\centerline{\includegraphics[scale=1.5]{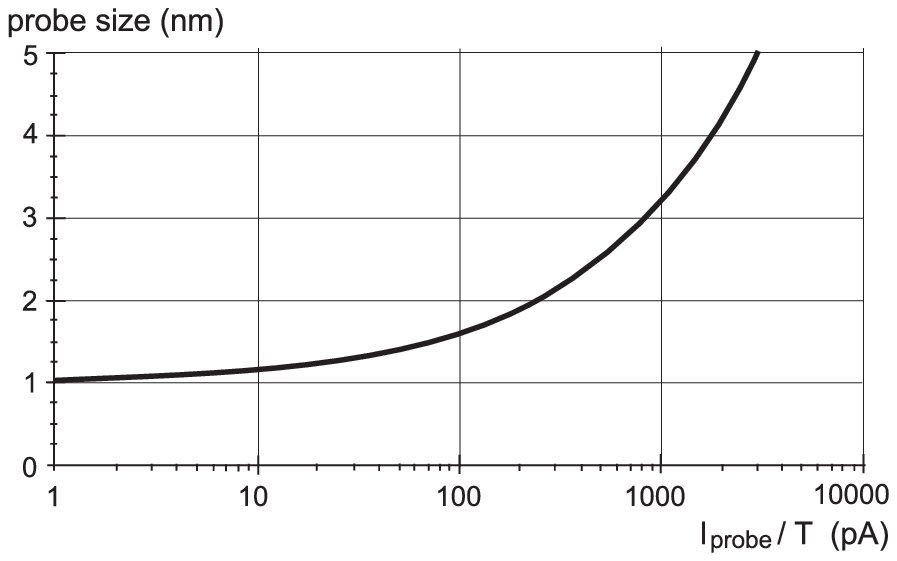}}
\label{fig11}
\end{figure}

$\phantom{a}$
\vspace*{-10mm}

\begin{center}
{Fig. 15. Probe size versus the probe current divided by
transmission ratio of~current through the foil}
\end{center}

$\phantom{a}$
\vspace*{-4mm}

The optimal semi convergent angle is 0,021--0,026~rad, and WD $ \approx $
1,8~mm.

The above examples clearly show advantages originating from thin film
application.
However, for practical realization of~a~SEM, and even more so array of
microcolumns, a~lot of~improvements has to be done yet. That concerns both
construction of~electron optical components and technology process for their
embodiment.

\vskip15pt

\begin{center}
{\large{\textbf{The detectors}}}
\end{center}

\vskip8pt

For transforming a~microcolumn into SEM is necessary to equip it with
appropriate detector of~secondary and back scattered electrons (SE and BSE)
as it sketched in fig.\:2.

Besides small size the SE and BSE detector has to fulfill all common
requirements: high collection efficiency, high gain at low voltages, fast
response time and linearity in wide range of~beam current. The most
promising contender seems to be micro channel plate (MCP) and pin diode
connected in a~tandem manner. Particular construction of~detector and
technology process for its manufacturing and assembling with microcolumn has
to be developed.

\vskip15pt

\begin{center}
{\large{\textbf{The technology for microcolumn manufacturing}}}
\end{center}

\vskip8pt

So far for microcolumn manufacturing were used or hybrid technology or MEMS
technology. Each of~those approaches has merits and demerits, which are well
known to those skilled in the art, so it is not discussed here in details.
We have being developing a~technology similar to that in micro electronics,
which seems to be more suitable for mass production. Have been developed
technology process for lenses shown in fig.\:8. (see fig.\:16 and
17).

Also have been developed electrostatic octopole deflector--stigmator with
thickness of~electrodes about 10~$\mu $m, sketched in fig.\:18 and 19.
Is assumed to implement two identical pieces placed between lenses in order
to achieve deflection simultaneously with stigmation.
\vskip16pt

\begin{center}
{\large
\textbf{Conclusion}
}
\end{center}
\vskip8pt

Since 1990-th when first miniaturized lenses have been micromachined,
a~substantial progress has been achieved in both methods for analytical
computations of~performance micro electron optics and prototyping
of~individual electron optical elements. As for arrayed microcolumns, the only
example of~matrix $4\times 4$ microcolumns for lithography purposes was presented
by T.H.P.~Chang and alii at  2000-th.

Nevertheless, in the foreseeable future one can expect the appearance of~the
systems manufactured by more advanced technology.

By our opinion the progress in this area would be determined by technology
starting from manufacturing individual components and ending with assembling
complete system. So when developing technology for any element as cathode,
lens or detector is necessary to think from the beginning about its
compatibility with whole technology process.

$\phantom{a}$
\vspace*{-20mm}

\noindent

\begin{table}[htbp]
\begin{tabular}
{p{360pt}}
\centerline{\includegraphics[scale=2.3]{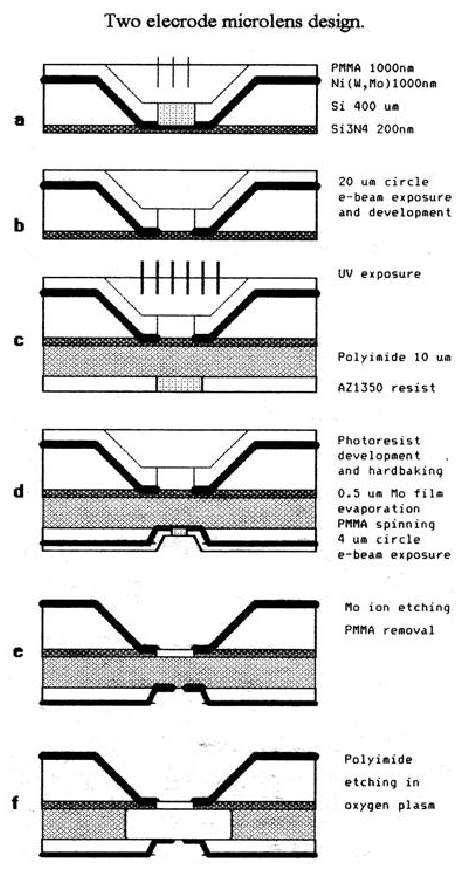}}\\
\label{fig11}
\centerline{Fig. 16. Non symmetrical two electrodes lens design}
\end{tabular}
\label{tab4}
\end{table}



\begin{table}[htbp]
\begin{tabular}
{p{360pt}}
\centerline{\includegraphics[scale=1.1]{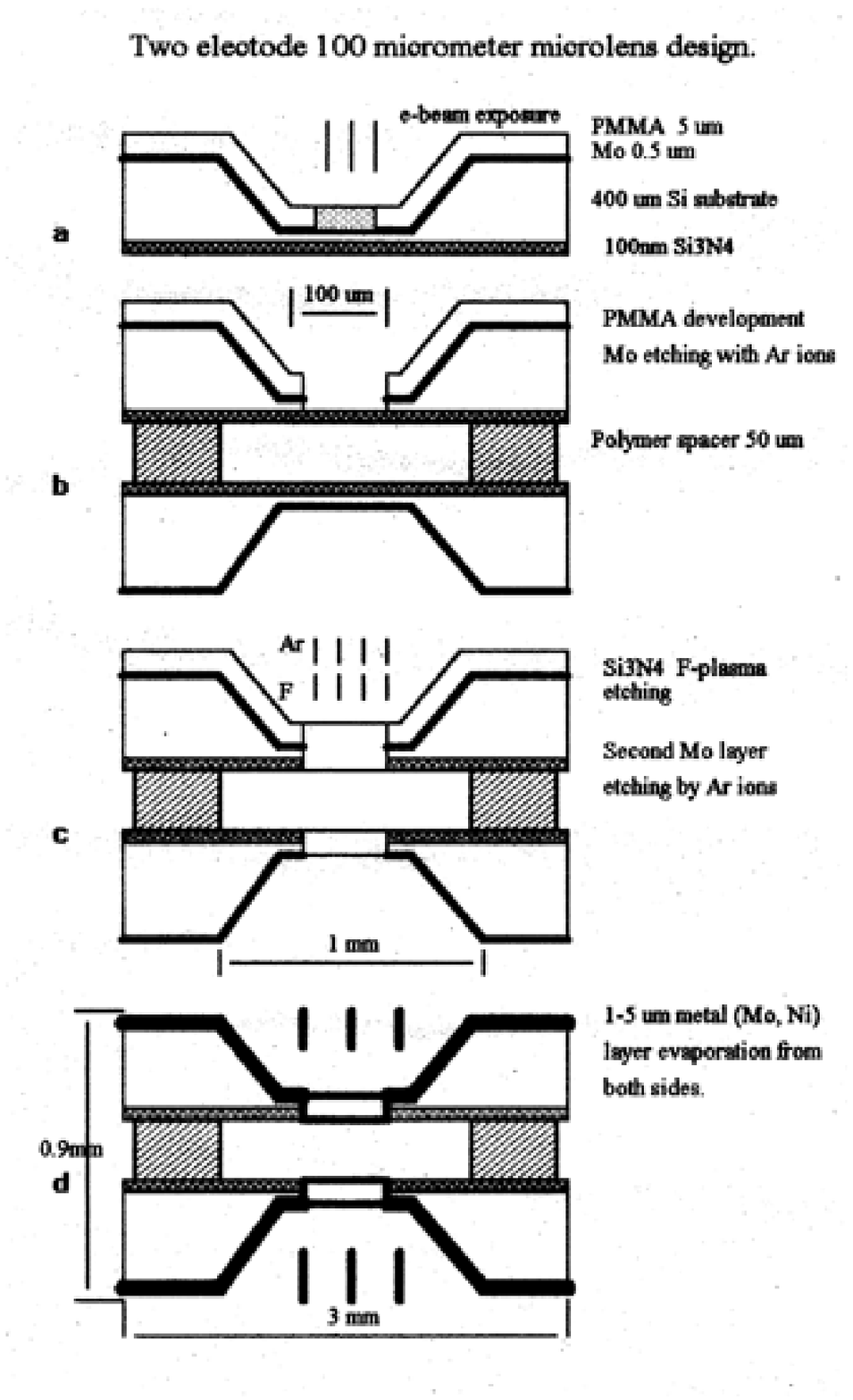}}\\[-10pt]
\centerline{Fig. 17. 100~$\mu$m micro lens design}
\end{tabular}
\label{tab4}
\end{table}



\par
\pagebreak

\begin{figure}[htbp]
\centerline{\includegraphics[scale=1.0]{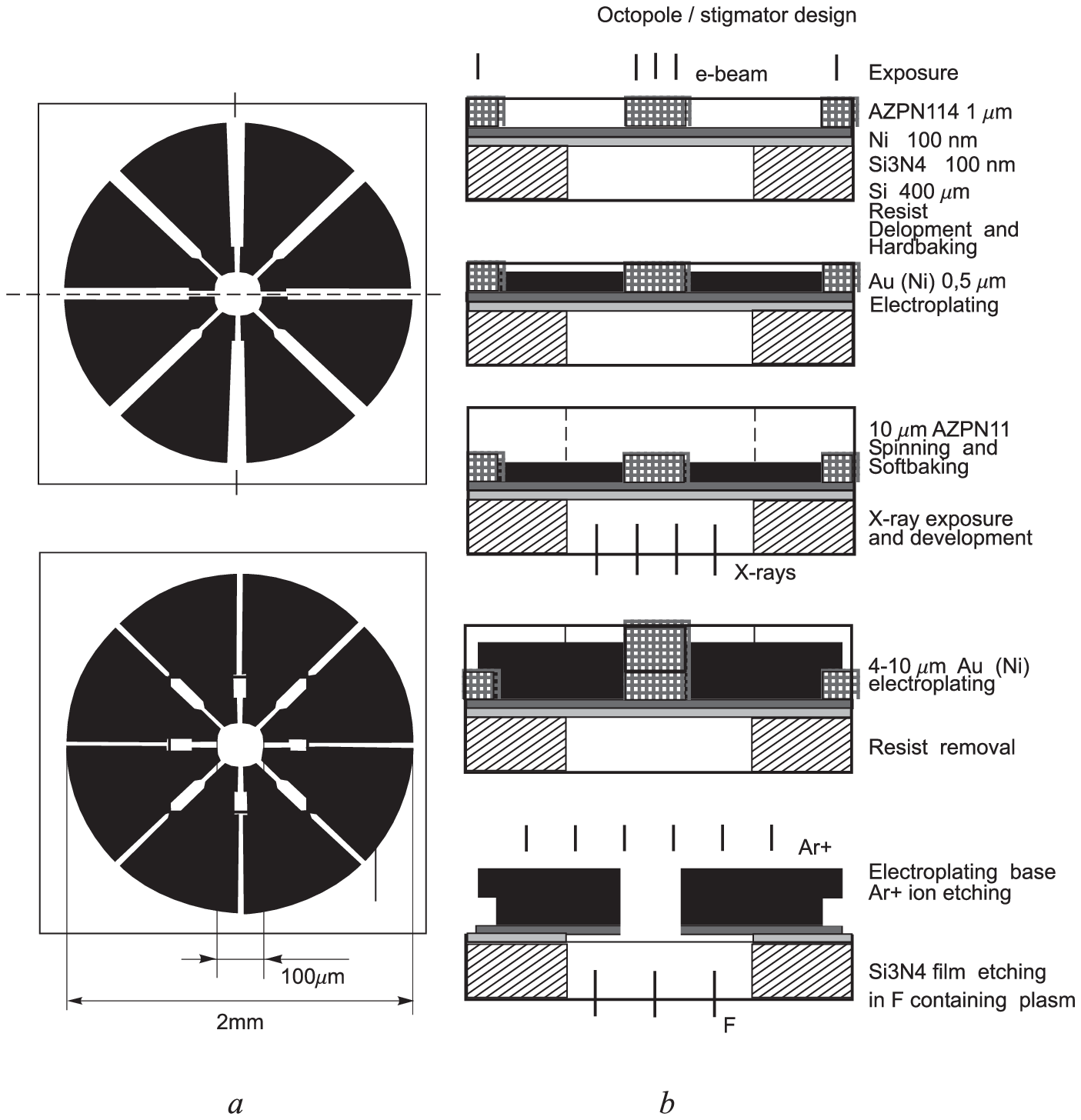}}
\label{fig11}
\end{figure}

\centerline{Fig. 18. Design of~octopole stigmator. Top view (\textit{a})
and axial section (\textit{b})}

\vskip40pt


{\large
{\begin{center}
\textbf{References}
\end{center}
}}
\vskip8pt

\begin{itemize}
\item[\hbox{[1]}]\textit{T.H.P. Chang, M.G.R. Thompson, M.L. Yu,
E. Kratshmer, et all}. Electron beam
microcolumn technology and application // SPIE. Vol.\:2522. P.\:4--12, (invited paper).

\item[\hbox{[2]}]\textit{G.P.E.M. van Bakel, E.G. Borgonjen,
C.W. Hagen, and P. Kruit}. Calculation of~the electron-optical
characteristics of~electron beams transmitted into
vacuum from a~sharp tip-thin foil junction // J. of~Appl. Phys. Vol.\:83
(1998) P.\:4279.

\item[\hbox{[3]}]\textit{R.H. van Aken, M.A.P.M. Janssen,
C.W. Hagen and P. Kruit}.  A simple fabrication method for tunnel
junction emitters // Solid State Electronics. Vol.\:45. (2001) P.\:1033.

\item[\hbox{[4]}]\textit{R.H. van Aken, C.W. Hagen,
J.E. Barth and P. Kruit}. Low-energy foil aberration
corrector // Ultramicroscopy. Vol.\:93. (2002) P.\:321.

\item[\hbox{[5]}]\textit{R.H. van Aken, C.W. Hagen,
J.E. Barth and P. Kruit}. Design of~a~low-voltage SEM
equipped with the low-energy foil corrector //
Submitted to Ultramicroscopy.

\item[\hbox{[6]}]\textit{V.V. Aristov, V.V. Kazmiruk,
V.A. Kudryashov, V.I. Levashov, S.I. Red'kin, C.W. Hagen, and P. Kruit}.
Microfabrication of~ultrathin free-standing platinum foils // Surface
Science.  V.\:337. 1998. P.\:402--404.

\item[\hbox{[7]}]\textit{H. S. Kim, S. Ahn, D. W. Kim,
Y. C. Kim and H. W. Kim, S. J. Ahn}. Sub-60-nm Lithography
Patterns by Low-Energy Microcolumn Lithography // Journal
of~the Korean Physical Society. Vol.\:49.  P.\:S712--S715.

\item[\hbox{[8]}]\textit{Ho-Seob Kim, Dae-Wook Kim, Seungjoon Ahn,
Sung-Soon Park, Myeong-Heon Seol and Young Chul Kim,
Sang-Kook Choi and Dae-Yong Kim}. Multi-Beam Microcolumns
Based on Arrayed SCM and WCM // J. of~the Korean Physical Soc.
Vol.\:45,  No.\:5.  P.\:1214--1217.

\item[\hbox{[9]}]\textit{V.V. Kazmiruk, T.N. Savitskaja}.
arXiv:0805.0248v1 [physics.ins-det].

\item[\hbox{[10]}]\textit{J.E. Barth}. Private communication.

\item[\hbox{[11]}]\textit{V.V. Dremov, V.A. Makarenk,
S.Y. Shapoval, O.V. Trofimov, V.G. Beshenkov
and I.I. Khodos}.  Sharp and clean Tungsten
Tips for STM investigation // Nanobiology. Vol.\:3. (1994) P.\:83--88.

\item[\hbox{[12]}]\textit{R.H. van Aken, M. Lenc and J.E. Barth}.
Aberration integrals for the low-voltage foil corrector // Nuclear
Instruments and Methods in Physics Research A. Vol.\:519. (2004)
P.\:205 and Vol.\:527 (2004) P.\:660.

\end{itemize}s

\end{document}